\documentclass[12pt]{article}
\usepackage{epsfig,amsmath}

\title{\bf A family of relativistic charged thin disks with an inner
edge}

\author{
{\sc Antonio C. Guti\'errez-Pi\~{n}eres}$^{1,2}$\thanks{e-mail:
agutierrez@ciencias.uis.edu.co, guillego@uis.edu.co, vivo1225@gmail.com}
\and
{\sc Guillermo A. Gonz\'{a}lez}$^{1,3}$\footnotemark[1]
\and
{\sc Viviana M. Vi\~{n}a-Cervantes}$^{1}$\footnotemark[1]\\
\newline
\and
$^{1}$Escuela de F\'{\i}sica, Universidad Industrial de Santander\\
A.A. 678, Bucaramanga, Colombia
\and
$^{2}$Facultad de Ciencias B\'asicas, Universidad de Santander\\
Bucaramanga, Colombia
\and
$^{3}$Departamento de F\'{\i}sica Te\'orica, Universidad del Pa\'is
Vasco\\
48080 Bilbao, Espa\~na}

\date{ }

\begin{document}

\maketitle

\begin{abstract}
A new family of exact solutions of the Einstein-Maxwell equations for static
axially symmetric spacetimes is presented. The metric functions of the solutions
are explicitely computed and are simply written in terms of the oblate
spheroidal coordinates. The solutions, obtained by applying the Ernst method of
complex potentials, describe an infinite family of static charged dust disks
with an inner edge. The energy density, pressure and charge density of all the
disks of the family are everywhere well behaved, in such a way that the
energy-momentum tensor fully agrees with all the energy conditions.\\

\end{abstract}

\section{Introduction}\label{sec:int}

The study of axially symmetric solutions of the Einstein and Einstein-Maxwell
field equations corresponding to disklike configurations of matter, apart from
its purely mathematical interest, has a clear astrophysical relevance. Now,
although normally it is considered that disks with electric fields do not have
clear astrophysical importance, there exists the posibility that some galaxies
be positively charged \cite{BH}, so the study of charged disks may be  of
interest not only in the context of exact solutions. On the other hand, the
observational data supporting the existence of black holes at the nucleus of
some galaxies is today so abundant, with the strongest dynamical evidence coming
from the center of the Milky Way, that there is no doubt about the relevance of
the study of binary systems composed by a thin disk surrounding a central black
hole (see \cite{CMS,BEG} for recent reviews on the observational evidence).

Now, as was shown by Lemos and Letelier \cite{LL1, LL2, LL3}, it is clear that
the obtaining of exact solutions that properly describe thin disklike sources
with an inner edge has a clear relevance in the study of binary systems that
involve a central black hole. Indeed, as was pointed in \cite{SEM4}, the main
properties of these annular disks strongly depend on their specific density
profiles. Therefore, it is worth to have other solutions, in addition to those
obtained previously, in such a way that a thoroughly study of such systems can
be made. Also, the obtaining of static solutions can be considered as the first
step in the obtaining of the more realistic stationary solutions, which can be
obtained from the previous ones by means of well known methods of solutions
generation.

In agreement with the above considerations, in this paper we present an infinite
family of new exact solutions of the Einstein-Maxwell equations for static
axially symmetric spacetimes. These solutions describe a family of relativistic
charged thin disks with an inner edge, the charged version of the family of
relativistic thin dust disks with an inner edge recently presented in
\cite{GG-PV1,GG-PV2}. The solutions were obtained by applying the Ernst method
of complex potentials \cite{Ern1,Ern2}, which generates a solution of the
Einstein-Maxwell equations starting from a solution of the vacuum Einstein
equations. The paper is organized as follows. First, in Section \ref{sec:einm},
we present the formalism to describe thin disks. Then, in Section
\ref{sec:discs}, we present a family of charged thin disks with an inner edge,
while in Section \ref{sec:beha} we analize the behavior of the solutions.
Finally, in Section \ref{sec:conc}, we conclude by summarizing our main results.

\section{Einstein-Maxwell Equations and Thin Disks}\label{sec:einm}

The vacuum Einstein-Maxwell equations, in geo\-me\-trized units such that $c =
8\pi G = \mu _{0} = \epsilon _{0} =  1$, can be written as 
\begin{eqnarray}
G_{ab} &=& T_{ab},  \label{eq:emep1} \\
&   &     \nonumber    \\
{F^{ab}}_{;b} &=& 0, \label{eq:emep2}
\end{eqnarray}
with the electromagnetic energy-mementum tensor given by
\begin{equation}
T_{ab} = F_{ac} F_b^{ \ c} - \frac{1}{4} g_{ab} F_{cd}
F^{cd} , \label{eq:emtensor}  
\end{equation}
where
\begin{equation}
F_{ab} =  A_{b,a} - A_{a,b} \label{eq:fab}
\end{equation}
is the electromagnetic field tensor and $A_a$ is the electromagnetic four
potential.

Now, for a static axially symmetric spacetime the line element can be written in
cylindrical coordinates $x^a = (t, \varphi, r, z)$ as \cite{KSMH}
\begin{eqnarray}
 \mathrm ds^2 = - \mathrm e^{2 \Phi} \mathrm dt^2 \ + \mathrm e^{- 2 \Phi}
 [r^2{\mathrm d}\varphi^2  + \mathrm e^{2 \Lambda} (\mathrm dr^2 + \mathrm
 dz^2)], \label{eq:met} 
\end{eqnarray}
where the metric functions $\Phi$ and $\Lambda$ depend on $r$ and $z$ only. We
take the electromagnetic potential as
\begin{equation}
A_a = (\psi, 0, 0, 0),
\end{equation}
where it is assumed that the electric potential $\psi$ also depends on $r$ and
$z$ only.

The solutions of the Einstein-Maxwell equations corresponding to a disklike
source are even functions of the $z$ coordinate. Therefore, they are everywhere
continuous functions but with their first $z$-derivatives discontinuous at the
disk surface. Accordingly, in order to obtain the energy-momentun tensor and the
current density of the source, we will express the jump across the disk of the
first $z$-derivatives of the metric tensor as 
\begin{equation}
 b_{ab} =  [{g_{ab,z}}] = 2 {g_{ab,z}}|_{_{z = 0^+}},
\end{equation}
and the jump across the disk of the electromagnetic field tensor as
\begin{equation}
[F_{za}] = [A_{a,z}] = 2 {A_{a,z}}|_{_{z = 0^+}},
\end{equation}
where the reflection symmetry of the functions with respect to $z = 0$ has
been used.

Then, by using the distributional approach \cite{PH,LICH,TAUB} or the junction
conditions on the extrinsic curvature of thin shells \cite{IS1,IS2,POI}, the
Einstein-Maxwell equations yield an energy-momentum tensor as
\begin{equation}
T^{ab} = T_+^{ab} \theta(z) + T_-^{ab} [1 - \theta(z)] + Q^{ab} \delta(z),
\label{eq:emtot}
\end{equation}
and a current density as
\begin{equation}
J^a = I^a \delta(z),   \label{eq:courrent}
\end{equation}
where $\theta(z)$ and $\delta (z)$ are, respectively, the Heaveside and Dirac
distributions with support on $z = 0$. Here $T_\pm^{ab}$ are the electromagnetic
energy-momentum tensors as defined by (\ref{eq:emtensor}) for the $z \geq 0$ and
$z \leq 0$ regions, respectively, whereas that
\begin{eqnarray}
Q^a_b = \frac{1}{2}\{b^{az}\delta^z_b - b^{zz}\delta^a_b + g^{az}b^z_b -
g^{zz}b^a_b + b^c_c (g^{zz}\delta^a_b - g^{az}\delta^z_b)\}.
\end{eqnarray}
gives the part of the energy-momentum tensor corresponding to the disk source, and 
\begin{equation}
I^a  =[F^{az}]
\end{equation}
is the contribution of the disk source to the current density. Now, the surface
energy-momentum tensor of the disk, $S_{ab}$, and the surface current density,
$j^a$, can be obtained through the relations
\begin{eqnarray}
S_{ab} \ &=& \ \int Q_{ab} \ \delta (z) \ ds_n \ = \ e^{ \Lambda - \Phi} \
Q_{ab} \ ,    \\
&   &           \nonumber      \\
j_a \ &= & \ \int I_a \ \delta (z) \ ds_n \ = \  e^{\Lambda - \Phi} I_a ,
\end{eqnarray}
where $ds_n = \sqrt{g_{zz}} \ dz$ is the physical measurement of length in
the direction normal to the disk. 

For the metric (\ref{eq:met}), the only non-zero component of $S^a_b$ and 
$j_0$ are
\begin{eqnarray}
S^0_0 &=& \ 2 e^{\Phi - \Lambda} \left\{ \Lambda,_z - \ 2 \Phi,_z\right\},
\label{eq:emt11}\\
&  & \nonumber \\
S^1_1 &=& \ 2 e^{\Phi - \Lambda} \Lambda,_z ,\label{eq:emt22}\\
&  & \nonumber \\
j_0 &=& \  -2e^{\Phi - \Lambda} \psi_{,z},
\end{eqnarray}
where all the quantities are evaluated at $z = 0^+$. Therefore, the surface
energy-momentum tensor of the disk and the surface current density of the disk
can be written as
\begin{eqnarray}
S^{ab} &=& \varepsilon V^a V^b, \\
&&	\nonumber	\\
j^a &=& \sigma V^a,
\end{eqnarray}
where
\begin{equation}
V^a = e^{-\Phi} (1, 0, 0, 0 ) 
\end{equation}
is the velocity vector of the matter distribution. 

\section{Charged dust disks with a central inner edge}\label{sec:discs}

In this section we consider a family of static axially symmetric solutions of the
vacuum Einstein-Maxwell equations given, for $n = 1,2,\dots$, by
\begin{eqnarray}
e^{\Phi_n} &=& \frac{2\beta e^{\phi_n}}{(1 + \beta) - e^{2\phi_n}(1 - \beta)},
\label{eq:Phin} \\
& & \nonumber\\
\psi_n &=& \frac{\sqrt{1 -\beta^2}(e^{2\phi_n}-1)}{e^{2\phi_n}(1 - \beta) - (1 +
\beta)}, \label{eq:psin} \\
&&	\nonumber	\\
\phi_n &=& \frac{\alpha y F_n(x,y)}{a^n(x^2 + y^2)^{2n -1}}, \label{eq:phin} \\
&&	\nonumber	\\
\Lambda_n &=& \frac{\alpha^2 (2n - 2)! (y^2 - 1) A_n (x,y)}{4^n a^{2n}
(x^2 + y^2)^{4n}}, \label{eq:lamn}
\end{eqnarray}
where $\alpha$ is an arbitrary constant and $\beta = \sqrt{1 - q^2}$, where $q^2
\leq 1$. The functions are all expresed in terms of the oblate spheroidal
coordinates, connected with the cylindrical coordinates via the transformation
formulas
\begin{align}\label{eq:oblates}
r^2 = a^2(1 + x^2)(1 - y^2),\quad z = axy,
\end{align}
where $x\in (-\infty,\infty)$ and $y\in[0,1]$. This solution was obtained, by
means of the Ernst method of complex potentials \cite{Ern1,Ern2}, from the
solutions describing a family of relativistic static thin dust disks with an
inner edge recently presented in \cite{GG-PV1,GG-PV2}.

In the above expressions, the functions $F_n (x,y)$ and $A_n (x,y)$ can be
easily obtained by means of the procedure described in \cite{GG-PV1}. So, the
functions $F_n (x,y)$ are polynomial functions with highest degree $4n-4$,  the
first two of them being
\begin{eqnarray}
F_1 &=& 1, \\
&& \nonumber \\
F_2 &=& x^4 + 3x^2 (1 - y^2) - y^2,
\end{eqnarray}
while the $A_n (x,y)$ are polynomial functions off highest
degree $8n-2$, of which the first two are
\begin{eqnarray}
A_1 &=& x^4 (9 y^2 - 1) + 2 x^2 y^2 (y^2 + 3) + y^4 (y^2 - 1), \\
&& \nonumber \\
A_2 &=& 2 x^{12} (9 y^2 - 1) - 4 x^{10} (51 y^4 - 41 y^2 + 2) \nonumber \\
&+& x^8 (735 y^6 - 1241 y^4 + 419 y^2 - 9) -x^6y^2 (132 y^6 \nonumber \\
&-& 1644 y^4 + 1604 y^2 - 252) + x^4 y^4 (84 y^6 - 384 y^4 \nonumber \\
&+& 1266y^2 - 630)+ 4 x^2 y^6 (6 y^6 + 6 y^4 - 39 y^2 + 63) \nonumber \\
&+& \ 3 y^8 (y^6 + y^4 + y^2 - 3),
\end{eqnarray}
all of them easily obtained as described in \cite{GG-PV1}.   

Now, the expressions for the surface energy density $\varepsilon$ and the
surface charge density $\sigma$ of the disks can be written as
\begin{eqnarray}
\varepsilon_n &=& 4 e^{\Phi_n - \Lambda_n}  \left\{1 - r \Phi_{n,r} \right\}
\Phi_{n,z} ,\\
&&	\nonumber	\\
\sigma_n &=& -2e^{\Phi_n - \Lambda_n}\psi_{n,z} , 
\end{eqnarray}
where all of the quantities are evaluated at $z = 0^+$. Accordingly, by using
the expressions (\ref{eq:Phin}) - (\ref{eq:lamn}), we obtain for the energy
density and the charge density the expressions
\begin{eqnarray}
\varepsilon_n &=& \frac{\epsilon_n}{\beta}, \label{eq:enern} \\
&&	\nonumber	\\
\sigma_n &=& \frac{q \epsilon_n}{2\beta}, \label{eq:charn}
\end{eqnarray}
where
\begin{equation}
\epsilon_n (x) = \frac{4 \alpha E_n (x)}{a^{n+1} x^{2n+1}}\exp \left\{ -
\frac{\alpha^2 (2n-2)! B_n (x)}{2^{2n} a^{2n} x^{4n}} \right\}, %\label{eq:enern}
\end{equation}
is the surface energy density of the static thin dust disks, without charge, as
presented in \cite{GG-PV1,GG-PV2}.

In the above expression, the $E_n(x)$ are positive definite polynomials of
degree $2k$, with $k = (n-1)/2$ for odd $n$ and $k = n/2$ for even $n$, of which
we only will write below the first three,
\begin{eqnarray}
E_1 (x) &=& 1, \\
&&	\nonumber	\\
E_2 (x) &=& x^2 + 3, \\
&&	\nonumber	\\
E_3 (x) &=& 3 (x^2 + 5),
\end{eqnarray}
and the $B_n (x)$ are positive definite polynomials of degree $4k$, with $k =
(n-1)/2$ for odd $n$ and $k = n/2$ for even $n$, being
\begin{eqnarray}
B_1 (x) &=& 1, \\
&&	\nonumber	\\
B_2 (x) &=& 2 x^4 + 8 x^2 + 9, \\
&&	\nonumber	\\
B_3 (x) &=& 27 x^4 + 72 x^2 + 50.
\end{eqnarray}
the first three of them.

\section{Behavior of the solutions}\label{sec:beha}

From the expressions in the previous section we can see that, by taking $\alpha
> 0$, the energy density of the disks will be everywhere positive,
\begin{equation}
\varepsilon_n (x) \geq 0.
\end{equation}
So that, as the azimuthal pressure is zero, we have an infinite family of dust
disks that are in fully agreement with all the energy conditions. Also is easy
to see that, for any value of $n$,
\begin{eqnarray}
\varepsilon_n (0) &=& 0, \\
&&	\nonumber	\\
\lim_{x \to \infty} \varepsilon_n (x) &=& 0.
\end{eqnarray}
That is, the energy density of the disks is zero at their inner edge and
vanishes at infinite. Furthermore, the surface mass density of the disks reduces
to their energy density,
\begin{equation}
\mu_n = \varepsilon_n, \label{eq:mun}
\end{equation}
so that its behavior is the same as of the energy density. Finally, by using
equation (\ref{eq:enern}) and (\ref{eq:charn}) we can write
\begin{eqnarray}
\sigma_n = \frac{q}{2} \varepsilon_n,
\end{eqnarray}
so that the charge density of the disks is equal, except by a constant, to their
energy density. Accordingly, the electric and gravitational forces are in exact
balance, as in the configurations of `electrically counterpoised dust'.

\begin{figure}
\begin{center}$$\begin{array}{r}
\epsfig{width=2.75in,file=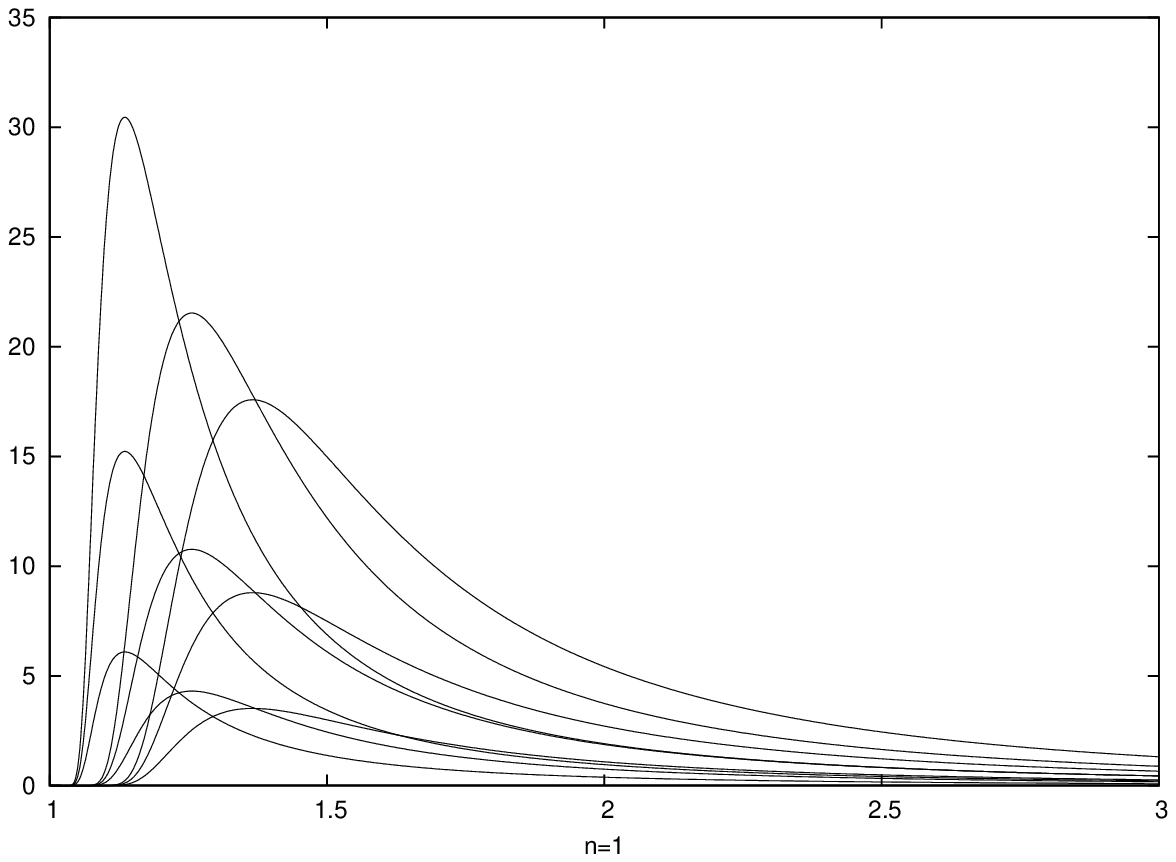} \\
\epsfig{width=2.75in,file=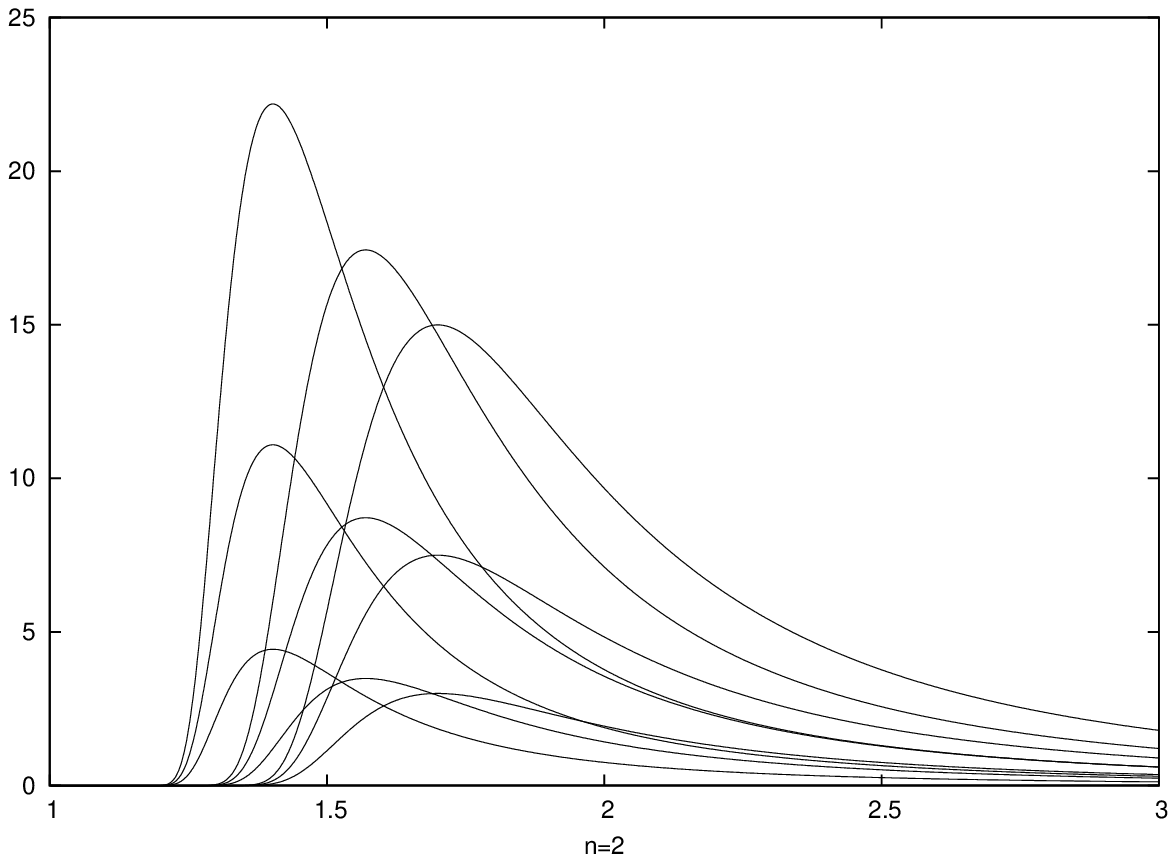} \\
\epsfig{width=2.75in,file=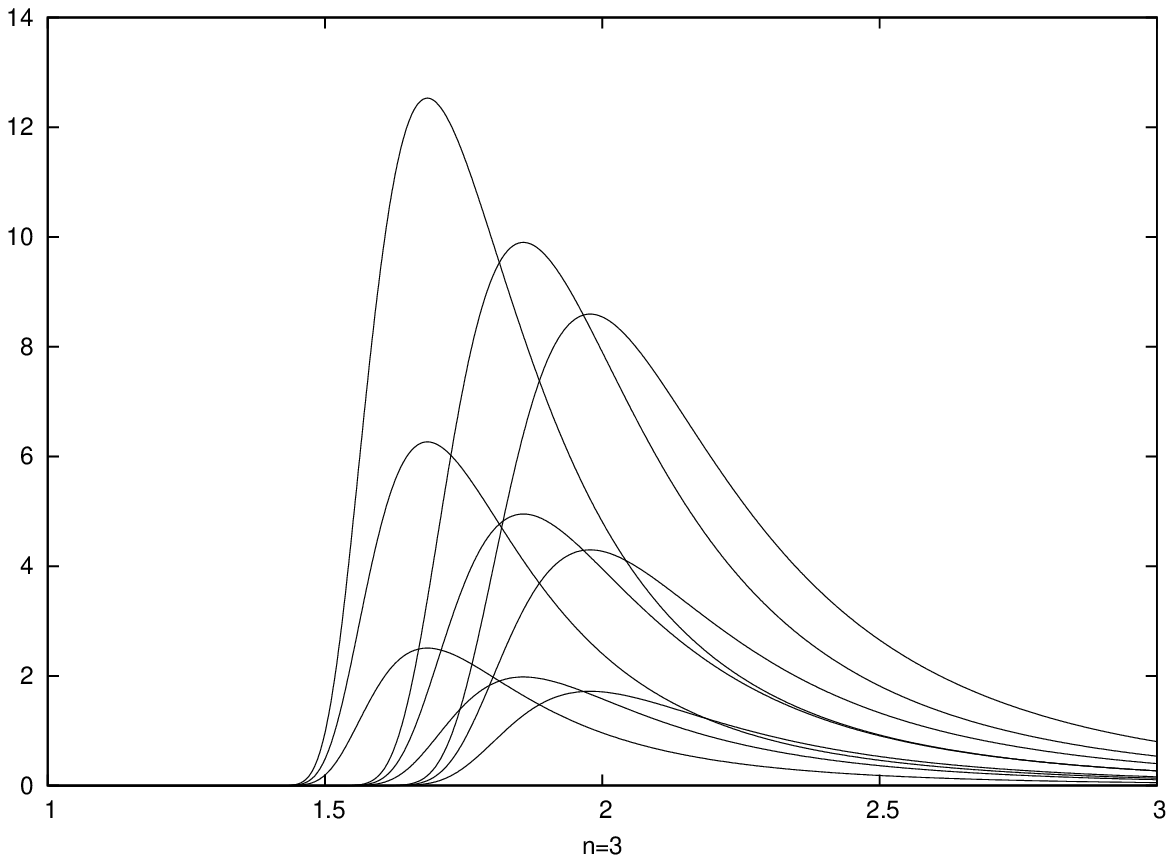} \\
\end{array}$$\end{center}
\caption{Energy density ${\tilde \varepsilon}_n$ as a
function of ${\tilde r}$ for the first three disks of the family, with ${\tilde
\alpha}_n = 0.5$, $1$, $1.5$ and $\beta=1$, $0.4$ and $0.2$. For each value of
$n$, the first curve on left corresponds to ${\tilde \alpha} = 0.5$, whereas
that the last curve on right corresponds to ${\tilde \alpha} = 4$. Also, for
each value of $n$ the curves with greatest maximum corresponds to $\beta=0.2$,
whereas the curves with least maximum corresponds to $\beta=1$, the disks
without charge}\label{fig:energy}
\end{figure}

Now, in order to show the behavior of the energy densities, we plot the
dimensionless surface energy densities ${\tilde \varepsilon}_n = a
\varepsilon_n$ as functions of the dimensionless radial coordinate ${\tilde r} =
r/a$. So, in Figure \ref{fig:energy}, we plot ${\tilde \epsilon}_n$ as a
function of ${\tilde r}$ for the first three disks of the family, with $n = 1$,
$2$ and $3$, for different values of the parameter ${\tilde \alpha}_n =
\alpha/a^n$. Then, for each value of $n$, we take ${\tilde \alpha}_n = 0.5$, $1$
and $1.5$.  The first curve on left corresponds to ${\tilde \alpha}_n = 0.5$,
while the last curve on right corresponds to ${\tilde \alpha}_n = 1.5$.  The
curves with greatest maximum corresponds to $\beta=0.2$, whereas the curves with
least maximum corresponds to $\beta=1$, the disks without charge. As we can see,
in all the cases the surface energy density is positive everywhere, having a
maximum near the inner edge of the disks, and then rapidly decreasing as
${\tilde r}$ increases. We can also see that, for a fixed value of $n$, as the
value of ${\tilde \alpha}_n$ increases, the value of the maximum diminishes and
moves towards increasing values of ${\tilde r}$. Also, as there is a direct
relationship between $\beta$ and the electric field, the maximum of the curves 
increases as the electric field increases.  The same behavior is observed for a
fixed value of ${\tilde \alpha}_n$ and increasing values of $n$.

\section{Concluding remarks}\label{sec:conc}

We presented an infinite family of new exact solutions of the vacuum
Einstein-Maxwell equations for static and axially symmetric spacetimes. The
solutions describe an infinite family of charged thin dust disks with a central
inner edge, the charged version of the family of relativistic thin dust disks
with an inner edge recently presented in \cite{GG-PV1,GG-PV2}. Now, although the
strange behavior of the Newtonian potentials may suggest that the disks do not
correspond to reasonable astrophysical sources, their energy densities are
everywhere positive and well behaved, in such a way that their energy-momentum
tensor are in fully agreement with all the energy conditions.

On the other hand, as all the metric functions of the solutions were explicitly
computed, these are the first fully integrated exact solutions for such kind of
thin disk sources. Moreover, the method used here to obtain these explicit
solutions may serve as a guideline to find more physical solutions in future
works. Now, besides their importance as a new family of exact solutions of the
Einstein-Maxwell vacuum equations, the main importance of this family of
solutions is that they can be easily superposed with the Schwarzschild solution
in order to describe binary systems composed by a thin disk surrounding a
charged central black hole, as was be done in \cite{GG1} with the first member
of the family of relativistic thin dust disks presented in
\cite{GG-PV1,GG-PV2}. 

\subsection*{Acknowledgments}

A. C. G-P. wants to thank the financial support from COLCIENCIAS, Colombia.

\end{document}